\documentclass[a4paper,12pt,double,oneside,openright,final]{article} 
\usepackage[english]{babel} 
\usepackage[utf8]{inputenc} 
\usepackage{calrsfs} 
\usepackage{amsmath} 
\usepackage{mathtools} 
\usepackage{breqn} 
\usepackage{amssymb} 
\usepackage{subeqnarray} 
\usepackage{caption} 
\usepackage{graphicx} 
\usepackage{epstopdf} 
\usepackage{pdfpages}
 \usepackage{cite}

\begin{document}
\title{Probing molecular environments with a fictitious isotopic dipole}
\author{José R. Mohallem and Paulo F. G. Velloso\\Laboratório de Átomos e Moléculas Especiais,\\ 
Departamento de Física, ICEx, Universidade Federal de Minas Gerais,\\ PO Box 702,
30123-970, Belo Horizonte, MG, Brazil\\ rachid@fisica.ufmg.br,\\\\Antonio F. C. Arapiraca\\Coordenação de Ciências,\\
Centro Federal de Educação Tecnológica de Minas Gerais,\\30421-169, Belo Horizonte, MG, Brazil}
\maketitle
\begin{abstract}
A HD-like isotopic dipole moment is proposed as a sensible probe for molecular environments, in particular for electrostatic fields and
 polarizable (reactive) sites of molecules. Fictitious nuclear masses are chosen in order to
yield a rigid dipole with appropriate magnitude. Upon subtracting the
Born-Oppenheimer energy, the interaction is reduced to the field-dipole-like and the dipole-polarizability-like terms, the last one being particularly informative since connected to potentially reactive sites. The field strength and orientation are 
easily obtained by identifying the minimum field-dipole energy configuration and flipping the dipole from it. In this case the method appears to have a superior accuracy in comparison with ab initio approaches. In tests with hydrogen, water, benzene and chlorobenzene
molecules and with a frustrated Lewis pair, the potential of the method is assessed.

\end{abstract} 

\section{Introduction}
In the last decades, the applicability of ab initio quantum chemical methods have been extended to the study of structural and dynamical properties of very large isolated molecules. Many important processes of modern science however, including those involving life, demand a step further, namely the generation of accurate theoretical knowledge of the properties of molecular environments, which are connected to the detailed description of Van der Waals (VdW) interactions and the identification of reactive sites for chemical processes \cite {Buck88,Maksic13}. The quite important topics of biological recognition \cite {Culver17}, hydrogen bonding \cite{Omeara15}, and computer simulations and modelling of molecular complexes and new materials \cite{Koc14,Wang14,Botu17}, for instance, lie on this subject. Particularly, the electrostatic field created by a source molecule on its surroundings is considered as being helpful for this prospect \cite{Koc14,Polit85,Gadre00,Kumar14,Mohan14,Sunda17}, since it indicates how the molecule affects statically its environment. But the knowledge of the molecular polarization  "potential", meaning the way the molecule would react dynamically to the presence of another, is of even greater importance \cite{Polit17}.

Reporting back to a review by Scrocco and Tomasi \cite{Scro78}, many investigations in these fields in the last decades rely on the molecular electrostatic potential (MEP) method, in order to investigate structure, reactivity and other properties of large molecules. Ab initio MEP derives directly from the electronic density \cite{Polit85,Polit17,Gadre00} and is the only method of general applicability so far. On the other hand, besides being static properties of isolated molecules, MEP fields are also inaccurate in regions close to the source molecule. This unsatisfactory situation motivated recent movements to the point charges model \cite{Cox81} and from the last to particular multipole expansions \cite{Pop14} and fragmented potentials \cite{Reid13}.  The situation is still unfavourable since, despite some particular tentatives \cite{Botu17,Pond02,Boeuf14,Gord17}, there is still no general approach to the real problem of predicting what happens when large molecules approach each other. 

We propose here a change of the  present paradigm to approach the problem, by exploring molecular environments with a computational probe that, simultaneously, evaluates the electrostatic field and  slightly interacts with the source molecule, identifying polarizable (perhaps reactive) sites.

VdW interactions are usually described by the well known classical fields \cite{Hozba88}. The partitioning of the quantum mechanical (VdW and stronger) energy in identifiable terms is not a well defined task, however, since the effects are all mixed in quantum chemical calculations. 
A prospective probe would allow us to \textit{turn on} just particular parts of the interaction energy and extract valuable information about the interaction, by isolating and identifying them. HD-like (D: Deuterium) isotopic dipole moments arise as good candidates. Besides HD being a small, neutral and closed-shell molecule and having reasonably small polarizability, its dipole moment
displays the here proposed feature that the interaction can be isolated, to a good extent, as that of a classical permanent dipole moment interacting with the source molecule. This dipole, made rigid by allowing just its rotational and translational degrees of freedom, will slightly polarize the source and, once this effect is separated, will \textit{align} to the source field.
 
This possibility arises from our ability to perform molecular electronic calculations accounting for the finiteness of
the nuclear masses, the so-called FNMC (finite-nuclear-mass-corrections)
approach \cite{Moha04,Moha11}, which has proven to yield correct experimental trends for isotopic dipole moments \cite{Ara11,Ara14,Ara16}. The FNMC electronic hamiltonian is

\begin{equation}
H=-\sum\limits_{A}(\sum\limits_{i}P_{A}\frac{\nabla_{i}^{2}}{2M_{A}}
P_{A})+H_{BO},
\label{eq1}
\end{equation}
in which $H_{BO}$ is the usual clamped-nuclei Born-Oppenheimer (BO) hamiltonian, $M_{A}$ is the mass of a generic nucleus, 
$P_{A}\frac{\nabla_{i}^{2}}{2M_{A}}P_{A}$ is the correction to the kinetic energy of electron $i$ due to the finiteness of
$M_{A}$ and $P_{A}$ projects the molecular electronic wavefunction on the
space of the atomic wavefunctions. The signature of the nuclear masses in hamiltonian (\ref{eq1}) allows to account for the isotopic effects already on the electronic level.

For the present purposes, the FNMC electronic energy of the probe-molecule system will be written as
\begin{equation}
E=E_{BO}+E_{dm}=E_{BO}+E_{df}+E_{dp}+E_{d+m},  \
\label{eq2}
\end{equation}
where $E_{BO}$ is the BO energy of an equivalent calculation for the probe-molecule system, $E_{dm}$ is the interaction energy
of the probe dipole with the molecule, $E_{df}$ is the energy of the dipole in the molecular electrostatic field,
$E_{dp}$ is the dipole-polarization energy and $E_{d+m}$ is the constant FNMC contribution of the isolated molecule and probe.
A most important feature is that $E_{BO}$ must account for all lead, non-isotopic, interaction  energy terms so that $E_{dm}$ accounts only for
the energy terms of the rigid dipole interacting with the source molecule.
Despite not being an issue, we do not subtract the positive constant term $E_{d+m}$, since the approach is not restricted to 
methods having size-consistency in FNMC calculations. 

Upon subtracting the BO energy, $E_{dm}$ is written
\begin{equation}
E_{dm}=E-E_{BO}=E_{df}+E_{dp}=-\mu\epsilon cos\theta+E_{dp}+E_{d+m},  \
\label{eq3}
\end{equation}
in which $\theta$ is the angle between the field $\boldsymbol\epsilon$ and the isotopic dipole moment $\boldsymbol\mu$ of the probe. 
Eqn. (\ref{eq3}) will allow the identification of the two components of $E_{dm}$ by fitting procedures. 

In the case of a polar molecule, $E_{df}$ will be strongly dominant. Otherwise, in the limit of a polarizable point molecule, we will find useful the classical formula for $E_{dp}$, namely

\begin{equation}
E_{dp}=-\dfrac{\alpha\mu^2 (3cos^2\theta+1)}{2 r^6},  \
\label{eq30}
\end{equation}  
in which $\alpha$ is the isotropic polarizability. Near the molecule, the dependence of $E_{df}$ and $E_{dp}$ with $r$ (distance from a chosen point of the molecule to the center of the dipole) is hardly so simple, however.
On the other hand, we can take advantage of the constancy of the electrostatic field in a fixed point and the proportionality of $E_{df}$ with $cos\theta$ in order to separate the $\theta$ dependence of the two terms, by turning the probe around a chosen axis which passes by the fixed point (see Figure 1, for example).
In cases in which we can consider  the molecule as a point particle, $E_{dp}$ will be proportional to $(3cos^2\theta+1)$ as in Eqn. (\ref{eq30}) and the procedure is much easier. 
In general cases, the dependence of $E_{dm}$ on $\theta$ can be checked, since the same dependence for $E_{df}$ is known. Furthermore, when we explore the symmetry of the molecule and the rigidity of the  probe dipole, the $\theta$ dependence of $E_{dp}$ can agree with Eqn. (\ref{eq30}) in particular situations, even when the dependence with $r$ does not. These features greatly simplify the analysis of the interaction, allowing a \textit{dissection} of $E_{dm}$, with both the
evaluation of the electric field and the characterization of $E_{dp}$.  These are the two goals of this work, which turns out as a powerful tool for the analysis of molecular environments. 
 
Once $E_{df}$ is isolated, the electrostatic field can be easily obtained. In fact, the dipole 
probe  will \textit{align} to it in the configuration for which $E_{df}$ is a minimum, fixing then the field orientation. As for its magnitude, let us define
\begin{equation}
\delta(E_{df,\theta})=(E_{df,0})-(E_{df,\theta}), \
\label{eq4}
\end{equation}
where $(E_{df,\theta})$ is calculated for the dipole assuming an angle $\theta$ from the field direction. 
Flipping the dipole to the
counter-aligned configuration, $\theta=\pi$ rad, the energy difference
$\delta(E_{df,\pi})$ yields the electric field magnitude as,
\begin{equation}
\epsilon=\frac{\delta(E_{df,\pi})}{2\mu}. \
\label{eq5}
\end{equation} 
An stringent test of this procedure is the accuracy of a fit of $E_{df}$ to a cosine function. In general, a trial and error fitting of $E_{dm}$ data to Eqn. (\ref{eq3}) gives information about $E_{dp}$ without previous knowledge of properties (dipole moment, polarizabilities, etc.) of the source molecule. Cases in which $E_{df}$ is largely dominant (correspondent to highly polar molecules) are more favourable since $E_{df}\simeq{E_{dm}}$.  Analogously, $E_{df}$ can be zero (in specific points) or negligible, so that $E_{dp}\simeq{E_{dm}}$. 

\section{Probe calibration}
The theoretical dipole moment of HD itself, $\mu=8.5\times10^{-4}$ Debye, is too small for our objectives. Calculations with FNMC are not accurate enough \cite{Ara11}, in consequence, so that the $E_{dm}$ components would suffer from the same drawbacks, mainly $E_{dp}$, in which $\mu$ enters squared. 
On the other hand, we are not constrained to work with a real probe. The FNMC method admits the use of fictitious nuclear masses, or fictitious isotopologues. With $M=10,000$ and $m=50$ a.u. chosen, after some tests, as  masses of, respectively, the heavy and light fictitious nuclei, 
the dipole moment of the probe results as $\mu=0.086$ Debye, two orders of magnitude larger than for HD. With good correlation method (CI or modern DFT) and basis set (typically superior than cc-pVDZ), the three decimal figures converge. These mass choice becomes a good compromise 
between having the largest isotopic dipole moment and keeping the validity of the adiabatic approximation for the
probe. 

Only the rotational and translational degrees of freedom are allowed to the probe, meaning that its internuclear distance $R$ is fixed. Its rigidity is then gauged by noticing that its isotropic polarizability differs by less than 1\% from that of $H_2$ in full-CI calculations with aug-cc-pVDZ basis set.
Since most of the polarization effects on the whole probe-molecule system are accounted for in the non-isotopic BO energy, which is subtracted to yield $E_{dm}$, further polarization effects on the probe become negligible. 

The probe dipole points from M to m, because M is more effective in attracting electrons (the Bohr radius of a one-electron M atom is smaller than that of a m atom), that is, M corresponds to the minus sign and m to the plus sign of the rigid dipole. 
As a final point concerning the probe calibration, in order the procedure of subtracting the BO energy to be consistent, the  length of  the probe is fixed at the equilibrium distance of $H_2$, $R=1.40$ a.u.

\section{Tests with H$_{2}$, H$_{2}$O, benzene and chlorobenzene}
Some tests with simple systems allow to gauge the performance of the probe. Besides being the smallest neutral molecule, $H_{2}$ has only an electric quadrupole moment, so it becomes an interesting test for the probe. 
At large distances, $r\geq8$ a.u., it behaves like a point molecule and the interaction energy $E_{dm}$ fits well a typical quadrupole $r^{-4}$ function, but its behaviour at shorter distances is more complicated as expected, see Figure 1 (at very short distances the $r^{-6}$ behaviour is dominant). 
The angular behaviour, on the other hand, is much simpler and more informative. At $r=5$ a.u., inset ($\textbf{a}$) displays $E_{dm}$ for full turning of the probe. Despite the complex behaviour with $r$ in this distance, the curve obtained has a very close angular behaviour to the classical dipole-polarizability $\theta$ dependence, $-(3cos^2\theta+1)$. We then use Eqn.  (\ref{eq30}) for the dipole-polarization term and fit $E_{dm}$ to Eqn.( \ref{eq3}) so that the isolation of $E_{df}$ yields a typical cosine function shown in inset ($\textbf{b}$), with energies of about $10^{-5}$ a.u.  The quadrupole field intensity is evaluated at that point as $\epsilon=2.6\times10^{-4}$ a.u. with rms deviation of $4\times10^{-6}$ for the cosine fit. The evaluation of such a small quadrupole field is indicative of the accuracy of the method.

Moving to the highly polar water molecule, we verify that the angular adjustment of $E_{dm}$ to a pure cosine function is very accurate in almost all the neighbourhood of the molecule, because $E_{df}$ is strongly dominant as expected, see Figure 2. However, approaching the lone electron pair of the oxygen atom along the symmetry molecular axis, we noted that from 7 to 5 a.u. from the this atom, the rms deviation of the fit increases by almost one order of magnitude (the deviation is also visible for 5 a.u.). This means that the $E_{dp}$ contribution increases and become progressively relevant. Fitting the results to Eqn. (\ref{eq3}) and isolating $E_{dp}$, we obtain the angular curve in the inset of Figure 2 at 5 a.u.  It shows a behaviour that we connect with a non-VdW region of the molecule, since $E_{dm}$ becomes sensible to direct and counter orientation of the probe. For
$\theta=\dfrac{\pi}{2}$, where the probe is aligned to the molecular axis, $E_{dp}$ has a small peak (m closer to the lone pair). At the counter aligned orientation the peak is higher (M closer to the lone pair, ) because of a larger repulsion. As we know, this is the region where hydrogen bonding occurs, so these features deserve further investigation.

The center of a benzene molecule, where there is no electric field, is another quite interesting site for checking the performance of the probe. The interaction is thus restricted to $E_{dp}$, but the point approximation for the molecule hardly applies, since we have various equivalent point atomic centers. In this case, both BO or FNMC energies show minima for the probe pointing to the middle of the bonds, as expected. Differently, $E_{dp}$ identifies the positions of the carbon atoms (the polarizable centers of the molecule) with minima, see Figure 3. The minimum of $E_{dm}=E_{dp}$ in the figure corresponds to the probe dipole pointing to the C$_{1}$ atom. 

A different pattern is displayed when we replace one of the H atoms by a Cl atom (chlorobenzene), since a non-zero electric field appears at the center of the ring. Figure 4 displays the result for $E_{dm}$. Considering the Cl atom as a point particle, the $E_{dp}$ classical contribution is subtracted from the $E_{dm}$ fit leading to an accurate cosine function for $E_{df}$, see the inset of Figure 4. We evaluate the intensity of the electric field as $\epsilon=5.3\times10^{-3}$ a.u., with $\boldsymbol\epsilon$ pointing to the Cl atom.

\section{Application to a frustrated Lewis pair}
Finally, we devise an application to a presently controversial problem involving larger molecules. A pair of an acid and a base molecules, sterically hindered by large substituents, is called a frustrated Lewis pair (FLP) \cite{Steph06,Steph08}. Such systems are not chemically neutralized and are able to activate molecular hydrogen and other smal molecules. 

Some theoretical models compete to explain this ability of FLPs. 
The simpler electric-field (EF) model \cite{Grim10} is based on the proposed existence of strong electric fields inside the FLP, mainly in regions close to the two central atoms, which would polarize the $H_2$ molecule and produce its bond cleavage. In \cite{Grim10}  it is suggested that the field intensity to accomplish this task should be larger than $0.1$ a.u., a huge electric field at molecular level.
In the previous static-DFT-based electron-transfer (ET) model, a more involved process of electron transfer from the FLP to $H_2$ and back to the FLP is suggested, weakening the $H_2$ bond so as to produce the cleavage \cite{Papai}. From DFT-MEP calculations, as a response to \cite{Grim10},  the authors contest the EF model by claiming that the electric field in the FLP cavity never reaches the predicted $0.1$ a.u value \cite{Rokob13}. However, it has been shown that DFT MEPS perform not better than Hartree-Fock MEPS, overestimating charge polarity and having particular problems in the presence of some atoms, including phosphorus \cite{Soli}, just the present case. More recent DFT metadynamics simulations calculations argue for an even more complex reaction mechanism, involving a series of transition states. So the correct mechanism is still a matter under debate.

Here we illustrate the potential of the present method by probing the cavity of the phosphane/borane FLP. We consider first some points around the medium point of
the line joining the central atoms P and B, in the minimum energy configuration of the FLP obtained withouth the $H_{2}$ molecule (from ref. \cite{Rokob13}). We use DFT with the B97D functional and the 6-31G** basis set for the molecule and the probe. 
As shown in the inset of Figure 5, the plot of $E_{dm}$ versus $\theta$ for point 2 is well adjusted by the two first terms of Eqn. (\ref{eq3}), with $E_{dp}$ given by Eqn. (\ref{eq30}). This means first that, one more time, the angular effect on the probe beyond the electrostatic $E_{df}$ is well approached by the classical dipole-polarization potential. This feature can be understood by realizing that the effect of the point central atoms must be largely dominant. Further, $E_{df}$ is not dominant but can be isolated and the electrostatic field evaluated.  
The components of $\boldsymbol\epsilon$ out of the P-B line result quite small. The \textit{moduli} of the electric field are evaluated in a.u. as $\epsilon_{1}=0.9\times10^{-2}$, $\epsilon_{2}=1.3\times10^{-2}$ and $\epsilon_{3}=2.1\times10^{-2}$. Despite further tests have shown an expected dependence ($20\%$ at most) of $\epsilon$ on both the DFT functionals basis sets, it is quite unlikely that it would reach a value near $0.1$ a.u. at any point, except perhaps in a physically unlike position much closer to the P atom. This means that our results do not support the assumption of too large electric fields of the EF model  \cite{Grim10} for $H_{2}$ activation as well. 
In view of this indication, we then consider the geometry corresponding to a transition state, the TS1 geometry of Liu et al \cite{Liu}, obtained in the presence of a $H_2$ molecule. Here the $H_2$ molecule is withdrawn and the probe is placed again in the middle point between the P and B atoms. Remarkably, the $E_{dp}$ contribution to $E_{dm}$ practically disappears, showing that the field in this position becomes purely electrostatic. This result is consistent with the shift of the $H_2$ molecule in TS1 from close to the middle of the central atoms to a position far from them \cite{Liu}, and with the corresponding transference of the FLP lone electron pair, more consistently with the ET model. Note however that it is obtained in the absence of the $H_2$ molecule, that is, it seems that the charge transfer is a configurational property of the FLP, independent of the presence of the molecule to be activated. 

\section{Conclusions}
In abstract, probing molecular environment represents a paradigmatic change as we compare with the limited MEP and the non general present approaches based on ab initio or semi-empirical methods. The fictitious isotopic probe proposed here is shown to work well in different environments, for the evaluation of electrostatic fields as well as the identification of polarizable, possibly reactive, sites in molecules. Also, the FNMC technique is quite easy to be upgraded to any method of quantum chemistry. It can thus become a powerful tool for the prediction of properties of large molecules, as illustrated with the last application to a FLP. 
\section{Acknowledgments}

JRM thanks Dr. Thomas Heine for calling his attention to the FLP problem. We thank Dr. Leonardo G. Diniz for his help in an earlier stage of this work and for useful discussions. This project is supported by CNPq  and Fapemig (Brazilian agencies).

\section{Figure captions}
FIGURE 1- Energy interaction of the probe with the $H_2$ molecule along its molecular axis. The triangles correspond to the calculated $E_{dm}$ and the full line to an adjustment of a $r^-4$ function from $r=8$ a.u.. Inset (a) shows $E_{dm}$ for $r=5$ a.u. and inset (b) shows the electrostatic quadrupole $E_{df}$.\\\\
FIGURE 2- Energy interaction of the probe with the water molecule along its axis and close to the O atom, for two distances. The inset shows $E_{dp}$ at $r=5$ a.u.\\\\
FIGURE 3- Energy interaction of the probe with the benzene molecule at its center.\\\\ 
FIGURE 4- Energy interaction of the probe with the chlorobenzene molecule at the center of the ring. The inset shows the electrostatic $E_{df}$.\\\\ 
FIGURE 5- The probe in the cavity of the FLP. The FLP drawing is only a pictorial representation not corresponding to any of the real configurations used here. The distances 1-2 and 2-3 are $0.44$ a.u. The inset shows the calculated points for $E_{dm}$ and their adjustment to Eqn. (\ref{eq3}) (see text for full explanation).


\end{document}